\documentclass[gmd, manuscript]{copernicus}

\usepackage{amsmath}
\usepackage{graphicx}
\graphicspath{{./}}
\usepackage{tikz}
\usepackage{tikz-3dplot}
\usetikzlibrary{shapes.geometric}

\definecolor{markC0}{HTML}{1F77B4}
\definecolor{markC1}{HTML}{FF7F0E}
\definecolor{markC2}{HTML}{2CA02C}
\definecolor{markC4}{HTML}{9467BD}

\usepackage{bm}
\usepackage{subcaption}
\usepackage{listings}
\usepackage{enumitem}
\usepackage{booktabs}
\usepackage{colortbl}
\usepackage{float}
\usepackage{pgfplotstable}
\pgfplotstableset{col sep=comma}
\usepackage{cleveref}
\crefname{lstlisting}{Listing}{Listings}
\Crefname{lstlisting}{Listing}{Listings}
\newcommand{\projectname}{TERRA-NG}

\newif\ifshowcomments
\showcommentsfalse

\definecolor{fbdarkgreen}{HTML}{006400}
\definecolor{bspurple}{HTML}{AE0080}
\ifshowcomments
  \newcommand{\nk}[1]{\textcolor{blue}{nk: #1}}
  \newcommand{\gr}[1]{\textcolor{magenta}{gr: #1}}
  \newcommand{\fb}[1]{\textcolor{fbdarkgreen}{fb: #1}}
  \newcommand{\mm}[1]{\textcolor{red}{mm: #1}}
  \newcommand{\bs}[1]{\textcolor{bspurple}{bs: #1}}
\else
  \newcommand{\nk}[1]{}
  \newcommand{\gr}[1]{}
  \newcommand{\fb}[1]{}
  \newcommand{\mm}[1]{}
  \newcommand{\bs}[1]{}

\fi

\begin{document}
\nolinenumbers 

\runningtitle{\projectname{} v1.0: Extreme-Scale, GPU-accelerated Mantle Convection}
\runningauthor{B\"ohm et al.}

\title{\projectname{} v1.0: Extreme-Scale, GPU-accelerated Mantle Convection}

\Author[1][fabian.boehm@fau.de]{Fabian}{B\"ohm}
\Author[3]{Nils}{Kohl}
\Author[3]{Ponsuganth}{Ilangovan}
\Author[3]{Gabriel}{Robl}
\Author[3]{Fatemeh}{Rezaei}
\Author[3]{Marcus}{Mohr}
\Author[3]{Bernhard S. A.}{Schuberth}
\Author[1,2]{Harald}{K\"ostler}
\Author[3]{Hans-Peter}{Bunge}
\Author[2,4]{Ulrich}{R\"ude}

\affil[1]{Erlangen National High Performance Computing Center (NHR@FAU), Erlangen, Germany}
\affil[2]{Chair for System Simulation (CS10), Friedrich--Alexander--Universit\"at Erlangen--N\"urnberg, Erlangen, Germany}
\affil[3]{Department of Earth and Environmental Sciences, LMU Munich, Germany}
\affil[4]{CERFACS, Toulouse, France}

\correspondence{Fabian B\"ohm (fabian.boehm@fau.de)}

\received{}
\pubdiscuss{}
\revised{}
\accepted{}
\published{}

\firstpage{1}

\maketitle

\begin{abstract}
We present \projectname{}, a portable, GPU-accelerated, matrix-free mantle-convection code. A single Kokkos C++ implementation runs at scale on NVIDIA, AMD, and Intel GPU supercomputers.  \projectname{} has a deliberately narrow design: built on a radially extruded mesh of spherical wedges, tailored to the spherical shell geometry, which enables domain-specific optimizations like single quadrature-point integral-evaluations, radial coordinate storage compression and radial shared-memory tiling. The corresponding low-order $W_1$-iso-$W_2/W_1$ wedge-based Stokes--energy discretisation is verified against the Zhong et al.~(2008) spherical-shell convection benchmark suite. We showcase TERRA-NG through strong- and weak-scaling on the JUWELS Booster (NVIDIA A100), MareNostrum 5 (NVIDIA H100), LUMI-G (AMD MI250X), Hunter (AMD MI300A APU), and SuperMUC-NG Phase 2 (Intel PVC) supercomputers.
Coupled mantle convection simulations at $\sim\!11$\,km and $\sim\!5.6$\,km radial spacing ($\sim 2.8$\,B and $\sim 22$\,B DoFs) can be run routinely on standard node partitions of all considered systems. Global $\sim\!1$\,km-per-gridpoint mantle convection ($\sim 1.4$\,T DoFs) is feasible on an extreme-scale allocation, and a sub-km hero-run at $\sim\!0.7$\,km grid spacing scaling up to $\sim 11,000$ GPUs of LUMI-G ($\sim 11$\,T\,DoFs) shows the potential of the code on future, larger machines.
\end{abstract}

\section*{Introduction}

The Earth's mantle is a roughly 3000\,km thick layer of rock between the Earth's
crust and the core and the largest part of the planet in terms of volume. Although stronger than steel on the time scale of hours,
it deforms and flows over geologic time. This creeping motion is known as
mantle convection. It is the driver of all large-scale tectonic activity on
our planet, including plate tectonics and mountain building, and
understanding it is essential to understanding the evolution of Earth and
other terrestrial planets~\citep{Davies-Book, schubert2001}.

Due to the high viscosity of mantle rock, on the order of $10^{21}$\,Pa\,s,
inertial forces are negligible relative to viscous forces, and the Reynolds
number of the flow is effectively zero. In this creeping-flow (Stokes)
regime the instantaneous velocity field is entirely determined by the
buoyancy forces and boundary conditions, and Rayleigh number and geometry
control the
pattern, wavelength, and amplitude of convective motion. This makes it
important to model mantle convection in its 3-D spherical-shell geometry. At
the same time, the mantle convects at very high Rayleigh number, resulting in
thin thermal boundary layers, narrow upwelling plumes and downwelling slabs,
and a broad range of interacting scales~\citep{Schuberth2009a,gassmoller2020,ilangovan2026}. 
High numerical resolution in 3-D spherical geometry is therefore essential 
for capturing the dynamics of the convecting mantle.

Early 3-D spherical mantle convection codes were developed in the mid-1980s,
with the pioneering work of \citet{baumgardner1985} and \citet{glatzmaier1988}.
The former introduced an icosahedral discretization
of the spherical shell~\citep{baumgardner1985icos} that provides a nearly
uniform resolution over the sphere. The resulting TERRA code was among the
first geodynamics codes to be parallelized for distributed-memory compute
clusters~\citep{bunge1995}, producing key results on the role of
depth-dependent viscosity and the resulting large-scale convective planform
of the mantle~\citep{bunge1996,bunge1997}.

A diverse ecosystem of mantle-convection codes has emerged over the past
four decades. Alongside TERRA, the community established CitcomS on a 12-cap
spherical mesh~\citep{zhong2000,zhong2008}, StagYY on the yin-yang
grid~\citep{tackley2008}, Underworld for particle-in-cell
rheology~\citep{moresi2007}, Fluidity for adaptive unstructured
discretizations~\citep{davies2011}, pTatin3D for matrix-free lithospheric
dynamics~\citep{may2014,may2015}, the octree-adaptive Rhea
code~\citep{burstedde2013}, ASPECT~\citep{kronbichler2012,heister2017},
which leverages the deal.II finite-element framework, and G-ADOPT~\citep{davies2022gadopt},
built on the automated finite-element system Firedrake~\citep{rathgeber2016firedrake}.
Stokes solves for mantle convection on pre-exascale CPU systems
reached six hundred billion degrees of freedom on 1.6
million cores in the 2015 Gordon Bell prize-winning work of
\citet{rudi2015}. TERRA itself was rebuilt on the HyTeG hierarchical hybrid
grid framework as part of the TerraNeo
project~\citep{kohl2019hyteg,bauer2019,bauer2020terraneo,ilangovan2026}.
GPU-targeted work for geodynamic Stokes problems has so far been more
limited in scope~\citep{zheng2014,kronbichler2019,clevenger2021,Lee:2026:EGUsphere}.

Despite such advances, the attainable resolutions remain below what is
needed to resolve the full range of dynamically active scales and the
interaction of convection with realistic, strongly temperature- and
stress-dependent rheologies.
For instance, the Earth's surface area is $4\pi R_o^2 \approx 5.1 \times 10^8$\,km$^2$ ($R_o = 6371$\,km). An assumed icosahedral surface grid at refinement level $\ell$ consisting of 10 diamonds holds roughly $N_\text{lat} = 10 \times 4^\ell$ nodes, yielding a lateral spacing of $h \approx \sqrt{4\pi R_o^2 / N_\text{lat}}$. At level 13, $N_\text{lat} \approx 6.7 \times 10^8$ and $h \approx 0.87$\,km. Matching this resolution radially across the 2891\,km deep shell requires $N_\text{rad} \approx 3300$ layers, giving $N = N_\text{rad} \cdot N_\text{lat} \approx 2.2 \times 10^{12}$ nodes and $\sim 1.1 \times 10^{13}$ degrees of freedom (DoFs;
assuming 5 unknowns per node: 3 velocity + 1 pressure + 1 temperature) for the momentum and energy balance.
A single solution vector at this resolution with double accuracy occupies $\sim 9 \times 10^{13}$ bytes ($\approx 88$\,TB).
 Storing a sparse matrix with a simple Laplacian stencil of $\sim$15 entries per row would require $\sim 2.6 \times 10^{15}$ bytes ($\approx 2.6$\,PB, 64-bit values and column indices),
 filling and exceeding the total GPU memory of contemporary top tier machines (\Cref{tab:sites}: LUMI-G aggregates $\sim 1.5$\,PB of HBM). Matrix-free methods~\citep{may2015,Kronbichler:2012:CAF}, which compute the effect of an operator application without assembling the associated global matrix,
are therefore essential. Furthermore, given the immense computational demands of the application, scalability and node-level performance are major concerns in mantle convection simulations. They are not an incremental upgrade for these simulations but the precondition that makes them feasible.

Simulations of this scale can only be conducted on supercomputers, where the hardware landscape has shifted decisively toward GPU-accelerated architectures, with all three major vendors competing: NVIDIA powers systems such as JUWELS Booster and JUPITER (Germany), MareNostrum 5 (Spain), Summit (USA), and ABCI (Japan); AMD equips Frontier and El Capitan (USA), LUMI-G (Finland), and Setonix (Australia); and Intel provides the GPUs for SuperMUC-NG Phase 2 (Germany) and Aurora (USA). The resulting heterogeneous landscape means that application codes must be portable across GPU backends to exploit the available resources.

\section*{Contribution}

In this paper, we present a new finite element code for 3-D spherical-shell
mantle convection designed from the outset for GPU architectures. By
exploiting the massive parallelism of modern GPUs, our code achieves
numerical resolutions substantially exceeding those of previous spherical
mantle convection models, opening the possibility of directly resolving
convective structures at a level of detail previously inaccessible to
global-scale simulations.
\projectname{} draws its numerical design from TERRA~\citep{baumgardner1985,baumgardner1985icos,bunge1995,Davies:2013:GMD}:
a single-application code for mantle convection on a radially extruded icosahedral mesh of spherical wedges, a geometry to which we tailor kernel-level optimisations that would not carry over to a general FE setting. Through Kokkos, a single source compiles unchanged across CUDA, HIP, SYCL, and multicore CPU backends and, paired with GPU-aware MPI, scales on NVIDIA, AMD, and Intel GPU supercomputers without per-vendor specialisation.

This paper presents the software architecture, the spherical-wedge Finite Element discretisation and its verification against spherical mantle convection
references, as well as a feasibility study of the coupled Stokes--energy simulation across five GPU supercomputers (\Cref{tab:sites}). In an extreme-scale access, \projectname{} reaches the unprecedented resolution of ${\sim}1$\,km per grid point in a global mantle-convection simulation for the first time, at timestep costs that would, given a sufficient allocation, permit such runs routinely. A companion paper~\citep{boehm2026performance} presents the cross-vendor node-level performance analysis of the compute kernels, which enabled reaching this resolution goal.

\bigskip

\noindent Throughout the paper, mesh resolution is parameterised by an integer level MT on the radially extruded icosahedral mesh, whose construction and data structure are described in detail in \Cref{sec:grid}: each diamond edge of the icosahedral shell carries MT lateral cells, and the shell is extruded through MT/2 radial layers, so e.g.\ MT128 corresponds to 128 lateral cells per edge and 64 radial layers. \Cref{tab:mt-resolution} lists the grid sizes, the implied Earth-mantle spacings, and the corresponding DoF counts for the $W_1$-iso-$W_2/W_1$ Stokes discretisation. The lateral spacing is $h_\mathrm{lat}$, tabulated at $R_\mathrm{surf}=6371$\,km and $R_\mathrm{CMB}=3480$\,km, and the radial spacing is $h_\mathrm{rad}$. The sweep spans a range from MT32 ($h_\mathrm{rad}\approx 181$\,km, $\sim 0.7$\,M DoFs, well coarser than any geodynamically meaningful flow model) to MT4096 ($h_\mathrm{rad}\approx 1.4$\,km, $\sim 1.4$\,T DoFs, at the resolution regime targeted in the introduction) up to sub 1km-scale with MT8192.

\begin{table}[H]
\centering
\small
\setlength{\tabcolsep}{5pt}
\caption{Per-macro-edge grid sizes, implied Earth-mantle resolution (assuming shell thickness $D=2891$\,km), and total DoF count for the $W_1$-iso-$W_2/W_1$ Stokes discretisation across the MT sweep. MT4096 is highlighted as the target of this work: it is the coarsest level reaching the kilometre-scale grid spacing that global mantle convection requires.}
\label{tab:mt-resolution}
\begin{tabular}{lccccc r}
\toprule
MT & lat & rad & $h^\mathrm{surf}_\mathrm{lat}$ & $h^\mathrm{CMB}_\mathrm{lat}$ & $h_\mathrm{rad}$ & DoF total \\
   & cells & layers & [km] & [km] & [km] &  \\
\midrule
MT32   & 32   & 16   & 220.4 & 120.4 & 180.7 & 0.72\,M \\
MT64   & 64   & 32   & 110.2 & 60.2  & 90.3  & 5.58\,M \\
MT128  & 128  & 64   & 55.1  & 30.1  & 45.2  & 43.9\,M \\
MT256  & 256  & 128  & 27.6  & 15.1  & 22.6  & 349\,M  \\
MT512  & 512  & 256  & 13.8  & 7.5   & 11.3  & 2.78\,B \\
MT1024 & 1024 & 512  & 6.9   & 3.8   & 5.6   & 22.2\,B \\
MT2048 & 2048 & 1024 & 3.4   & 1.9   & 2.8   & 177\,B  \\
\textbf{MT4096} & \textbf{4096} & \textbf{2048} & \textbf{1.7}   & \textbf{0.9}   & \textbf{1.4}   & \textbf{1.42\,T} \\
MT8192 & 8192 & 4096 & 0.85  & 0.46  & 0.71  & 11.4\,T \\
\bottomrule
\end{tabular}
\end{table}

\setcounter{section}{0}

\section{Model}
\label{sec:model}

Mantle convection is governed by the conservation of momentum (Stokes equations), conservation of mass (anelastic constraint), and an energy balance (advection--diffusion of temperature with adiabatic, shear and internal heating), coupled through thermal buoyancy and temperature-dependent rheology~\citep{schubert2001}. Mantle-convection models differ in how they treat compressibility, ranging from the Boussinesq approximation over the anelastic liquid approximation~\citep{gough1969,jarvisandmckenzie1980} to the fully compressible case, see \citet{gassmoller2020} for an overview of the formulations. Following the Truncated Anelastic Liquid Approximation (TALA) formulation~\citep{ilangovan2026}, the momentum and mass conservation equations read
\begin{align}
  -\nabla \cdot \boldsymbol{\tau} + \nabla p' &= \bar\rho\,\bar\alpha\,T'\,\hat{\mathbf{r}}, \label{eq:stokes}\\
  \nabla \cdot \bigl(\bar\rho\,\mathbf{u}\bigr) &= 0, \label{eq:mass}
\end{align}
with the deviatoric stress tensor
\begin{equation}
  \boldsymbol{\tau} \;=\; \eta\bigl(\nabla \mathbf{u} + \nabla \mathbf{u}^{\!\top}\bigr) - \tfrac{2}{3}\,\eta\,(\nabla\!\cdot\!\mathbf{u})\,\mathbf{I},
  \label{eq:stress}
\end{equation}
where $\mathbf{u}$ is the velocity, $p'$ the pressure perturbation from the adiabatic hydrostatic state, $\eta = \eta(\mathbf{x}, T)$ the viscosity, which varies spatially and with temperature, $\bar\rho = \bar\rho(r)$ and $\bar\alpha = \bar\alpha(r)$ the reference (adiabatic) density and thermal-expansivity profiles, $T'$ the temperature perturbation from the adiabatic reference, and $\hat{\mathbf{r}}$ the radial unit vector.

The energy equation governs the temperature field $T$:
\begin{flalign}
  & \frac{\partial T}{\partial t} + \mathbf{u}\!\cdot\!\nabla T \;-\; \frac{1}{\bar\rho\,\bar c_p}\,\nabla \cdot \bigl(k\,\nabla T\bigr)
  =\; \frac{1}{\bar\rho\,\bar c_p}\,\boldsymbol{\tau}\!:\!\dot{\boldsymbol{\varepsilon}} \;+\; \frac{\bar\alpha}{\bar c_p}\,(\mathbf{u}\!\cdot\!\mathbf{g})\,T \;+\; H,
  \label{eq:energy} &&
\end{flalign}
with $\bar c_p = \bar c_p(r)$ the reference specific heat, $k$ the thermal conductivity, $\mathbf{g}$ gravity, and the three RHS source terms accounting for shear heating, adiabatic heating, and internal (radiogenic) heating $H$. The two temperature fields are related by the additive split
\begin{equation}
  T(\mathbf{x},t) \;=\; \bar T(r) \;+\; T'(\mathbf{x},t),
  \label{eq:tsplit}
\end{equation}
where $\bar T = \bar T(r)$ is a static radial reference profile, taken either as the analytic steady-conduction solution or as a tabulated geotherm. The system is coupled: \eqref{eq:energy} is advanced in the full field $T$, only the deviation $T'$ enters the buoyancy term of \eqref{eq:stokes}, and the advection and adiabatic-heating terms in \eqref{eq:energy} depend on $\mathbf{u}$.

\section{Grid and Parallelization}
\label{sec:grid}

\begin{figure}[t]
\centering
\subcaptionbox{\centering Unfolded diamond layout\label{fig:diamonds}}[0.44\textwidth]
  {\resizebox{\linewidth}{!}{\begin{tikzpicture}[scale=0.34]

  \def\dw{1.4}
  \def\dh{2.0}
  \def\sep{3.0}
  \def\vgap{0.8}
  \def\xoff{0}

  \foreach \i in {0,...,4} {
    \pgfmathsetmacro{\cx}{\i*\sep + \xoff}
    \draw ({\cx},{\dh}) -- ({\cx+\dw},{0}) -- ({\cx},{-\dh}) -- ({\cx-\dw},{0}) -- cycle;
    \node[font=\small] at ({\cx},{0}) {\i};
  }

  \foreach \i in {0,...,4} {
    \pgfmathsetmacro{\cx}{\i*\sep + \sep/2 + \xoff}
    \pgfmathsetmacro{\cy}{-2*\dh+\vgap}
    \pgfmathsetmacro{\did}{int(\i+5)}
    \draw ({\cx},{\cy+\dh}) -- ({\cx+\dw},{\cy}) -- ({\cx},{\cy-\dh}) -- ({\cx-\dw},{\cy}) -- cycle;
    \node[font=\small] at ({\cx},{\cy}) {\did};
  }

  \draw[gray!70, dotted] ({-\sep/2+\xoff},{-2*\dh+\vgap+\dh}) -- ({-\sep/2+\dw+\xoff},{-2*\dh+\vgap}) -- ({-\sep/2+\xoff},{-2*\dh+\vgap-\dh}) -- ({-\sep/2-\dw+\xoff},{-2*\dh+\vgap}) -- cycle;
  \node[gray!70, font=\small] at ({-\sep/2+\xoff},{-2*\dh+\vgap}) {9};

  \draw[gray!70, dotted] ({5*\sep+\xoff},{\dh}) -- ({5*\sep+\dw+\xoff},{0}) -- ({5*\sep+\xoff},{-\dh}) -- ({5*\sep-\dw+\xoff},{0}) -- cycle;
  \node[gray!70, font=\small] at ({5*\sep+\xoff},{0}) {0};

\end{tikzpicture}}}
\hfill
\subcaptionbox{\centering Assembled shell\label{fig:shell:assembled}}[0.26\textwidth]
  {\includegraphics[height=2.6cm]{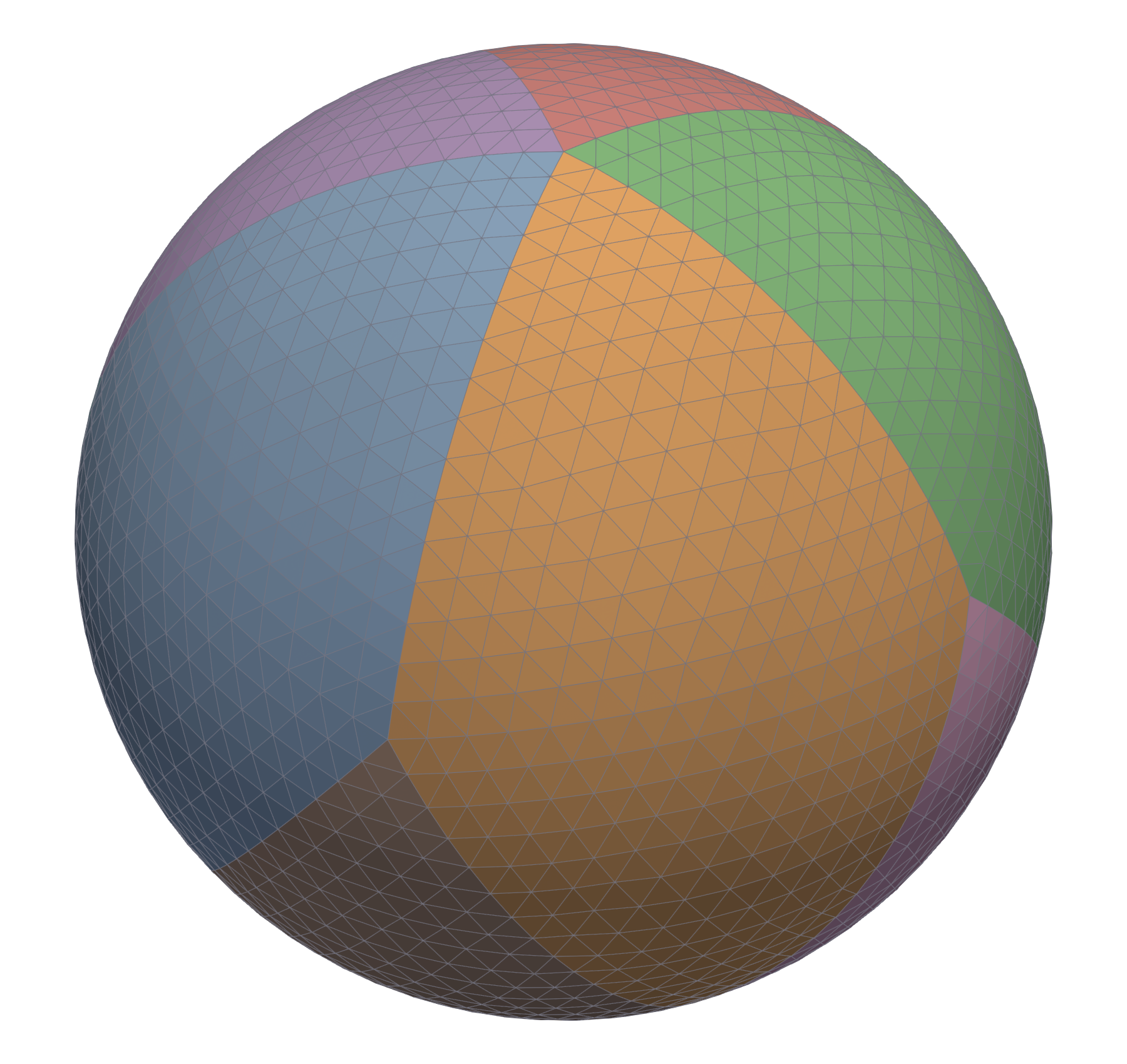}}
\hfill
\subcaptionbox{\centering Exploded view\label{fig:shell:exploded}}[0.26\textwidth]
  {\includegraphics[height=2.9cm]{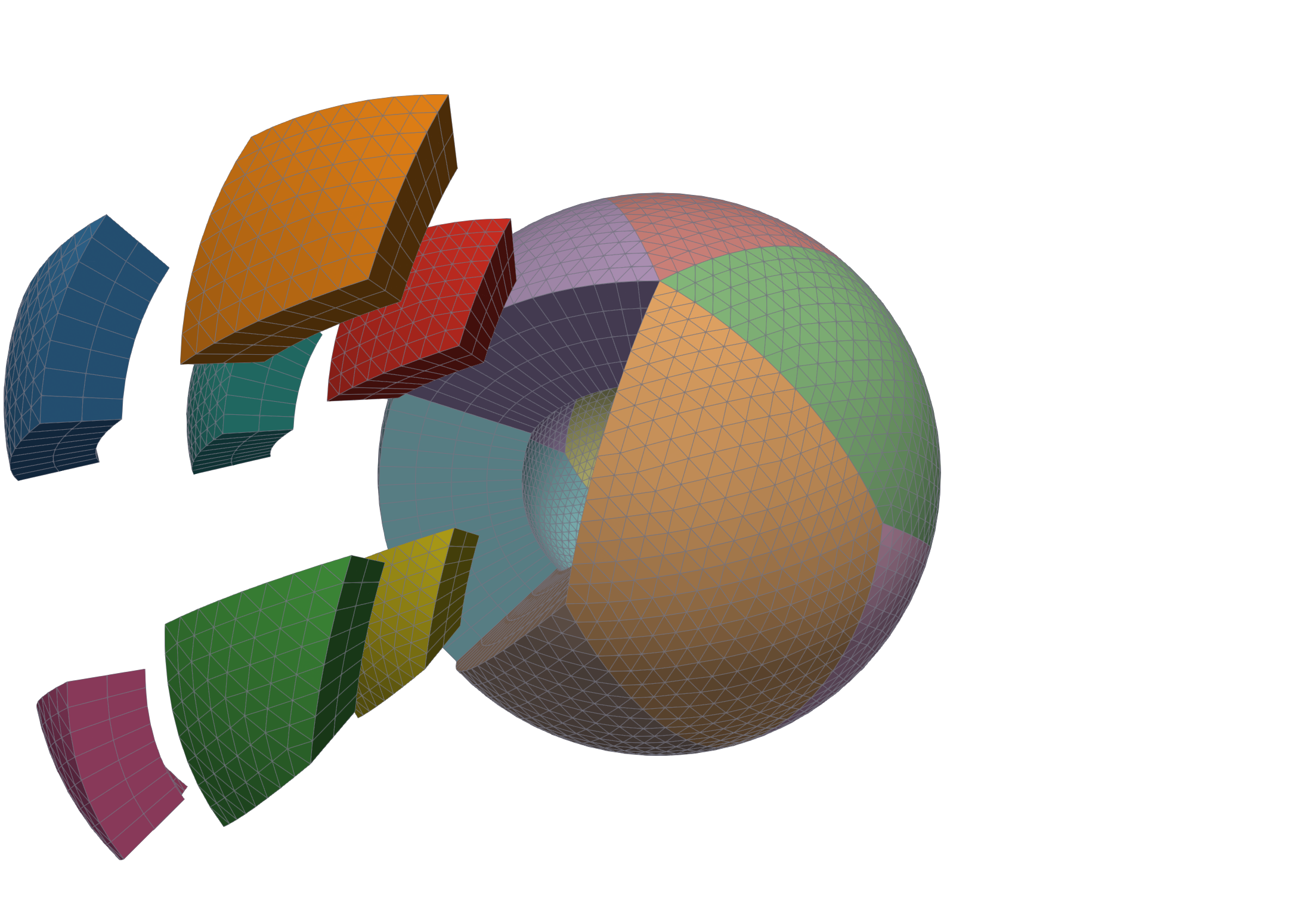}}
\caption{(a)~Unfolded layout of the 10 curvilinear diamonds with IDs 0--9. Ghost diamonds indicate periodic connectivity. (b)~3D Icosahedral spherical shell grid with the 10 diamonds highlighted in colors. (c)~Exploded view of a single diamond partitioned into 8 subdomains, each assigned to one MPI rank. An MPI rank can also hold multiple subdomains.}
\label{fig:shell}
\end{figure}
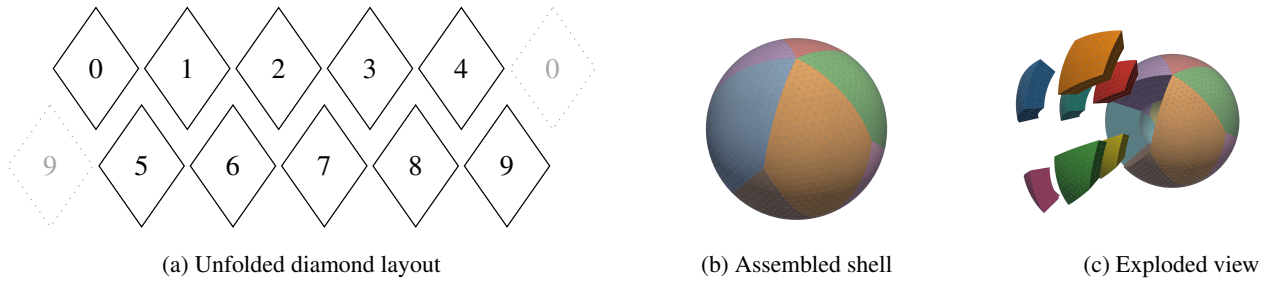

The spherical shell is discretized in two stages. First, the sphere's surface is partitioned into 10 diamonds composed of two spherical triangles based on an
icosahedral base grid with 12 vertices (\cref{fig:diamonds}). Each diamond is a curved quadrilateral spanned by 4 icosahedral nodes connected by geodesic arcs.
It is refined laterally by recursive midpoint refinement of the underlying
spherical triangles \citep{baumgardner1985icos}, yielding $2^\ell \times 2^\ell$ quadrilateral cells per diamond at refinement level $\ell$. Second, the surface mesh is extruded radially through $N_\text{rad}$ layers to fill the shell volume, with flexible spacing (\cref{fig:shell}).
Each hexahedral cell produced by the radial extrusion is split diagonally into two wedge (prism) elements. The splitting is implicit in the compute kernels and not stored in the grid data structure.

For parallel execution, each diamond block is partitioned into subdomains (\cref{fig:shell:exploded}). Each MPI rank holds one or more subdomains. The unknowns held by an MPI rank are stored in a \texttt{Kokkos::View} with dimensions $(n_{sd}, n_x, n_y, n_r)$, where $n_{sd}$ is the number of subdomains, $n_x$ and $n_y$ are the number of cells from refinement in the lateral directions and $n_r$ the number of radial layers. This rank-local data structure exposes a regular 4D index space that is parallelized across GPU threads. At compile time, the View's memory space is bound to the selected Kokkos backend, so the same declaration maps to \texttt{cudaMalloc} on NVIDIA GPUs, \texttt{hipMalloc} on AMD, or \texttt{syclMallocDevice} on Intel, without changing application code. The unknowns are transferred to GPU memory once at initialization and only copied back to the host for I/O. During the solve, all computation stays on the GPU; after each operator apply, boundary DoFs shared between neighbouring subdomains are exchanged using non-blocking, GPU-aware MPI on device-resident send and recv buffers, with packing and atomic-add unpacking implemented as Kokkos kernels in the same backend as the apply itself, avoiding any host staging. \Cref{lst:kokkos-view} shows the corresponding type aliases.

\begin{lstlisting}[
  caption={Grid data type aliases. The \texttt{ScalarType} template parameter enables mixed-precision instantiation. The \texttt{Grid4DDataVec} wrapper stores each vector component in a separate View (structure-of-arrays) for coalesced GPU memory access.},
  label={lst:kokkos-view},
  float=t,
]
using Layout = Kokkos::LayoutRight;

// Scalar grid data (4D: subdomain x lat_x x lat_y x radial)
template <typename ScalarType>
using Grid4DDataScalar = Kokkos::View<ScalarType****, Layout>;

// Vector grid data: structure-of-arrays for coalescing
template <typename ScalarType, int VecDim>
struct Grid4DDataVec {
  Grid4DDataScalar<ScalarType> comp_[VecDim];
};
\end{lstlisting}

\section{Spatial Discretization}
\label{sec:discretization}

The Stokes system \eqref{eq:stokes}--\eqref{eq:mass} is discretized with the $W_1$-iso-$W_2$ / $W_1$ element pair (velocity $W_1$-iso-$W_2$ over pressure $W_1$), where $W_k$ denotes the wedge element space spanned by the tensor product of a degree-$k$ triangular and a degree-$k$ radial basis, which satisfies the inf-sup (LBB) stability condition: velocity $\bm{u} \in [W_1]^3$ lives on the refined grid at level $\ell$ with 6 nodes per wedge (18 DoFs per element), while pressure $p \in W_1$ lives on the coarser grid at level $\ell - 1$. Both use the same tensor-product wedge basis functions on different grid levels: on the reference wedge with lateral coordinates $\xi, \eta \geq 0$, $\xi + \eta \leq 1$ and radial coordinate $\zeta \in [-1, 1]$, the six nodal basis functions are the products of $P_1$ triangle and $P_1$ radial bases,
\begin{equation}
  N_j \;=\; N^{\mathrm{lat}}_{j \bmod 3}(\xi,\eta)\; N^{\mathrm{rad}}_{\lfloor j/3 \rfloor}(\zeta), \qquad j = 0,\dots,5,
  \label{eq:wedgebasis}
\end{equation}
with $N^{\mathrm{lat}}_0 = 1{-}\xi{-}\eta$, $N^{\mathrm{lat}}_1 = \xi$, $N^{\mathrm{lat}}_2 = \eta$ and $N^{\mathrm{rad}}_0 = \tfrac12(1{-}\zeta)$, $N^{\mathrm{rad}}_1 = \tfrac12(1{+}\zeta)$, associating nodes $0$--$2$ with the bottom and $3$--$5$ with the top triangular face.

The element integrals are evaluated with a single quadrature point per wedge (two points per hexahedral cell); the accuracy and stability of this choice are analysed in \Cref{sec:quadrature}.
 The discretization is conforming: nodal degrees of freedom on element boundaries are shared between adjacent wedges. This synergises well with shared-memory tile staging.

As motivated in the introduction, storing the global sparse matrix is infeasible at the target resolution. Instead, the global matrix-vector product $\bm{y} = A\,\bm{x}$ is computed by summing element-local contributions on-the-fly, for example for the viscous part of \cref{eq:stokes}:
\begin{equation}
  \bm{y} = A\,\bm{x} = \sum_{e} I_e^\top\, A_e\, I_e\, \bm{x},
  \label{eq:matfree}
\end{equation}
where $I_e$ is the local-to-global mapping (gather) and $A_e$ is the element matrix arising from the weak form:
\begin{equation}
  (A_e)_{ij} = \int_{\Omega_e} \eta \Big[ 2\,\bm{\varepsilon}(\bm{\varphi}_i) : \bm{\varepsilon}(\bm{\varphi}_j) - \tfrac{2}{3}\,\mathrm{div}(\bm{\varphi}_i)\,\mathrm{div}(\bm{\varphi}_j) \Big] \,\mathrm{d}x,
  \label{eq:elemmat}
\end{equation}
where $\bm{\varphi}_i$ are the local basis functions. Each element contribution is computed independently: the local DoFs $\bm{x}_e = I_e\,\bm{x}$ are gathered from the global source vector, $A_e$ is assembled on-the-fly from geometry and basis functions, the local product $\bm{y}_e = A_e\,\bm{x}_e$ is computed, and the result is scattered back via atomic addition ($I_e^\top$). This avoids storing the global matrix entirely, reducing the memory footprint from $\mathcal{O}(N \cdot \mathrm{nnz/row})$ to $\mathcal{O}(N)$ at the cost of recomputing element matrices at every operator application. The details of the kernel implementation, key to achieving high throughput and spatial resolution in our simulations, are explained in the companion performance paper~\citep{boehm2026performance}.

\section{Single-Point Quadrature and Hourglass Modes}
\label{sec:quadrature}

The wedge basis functions~\eqref{eq:wedgebasis} span the six-dimensional
local space
\begin{equation*}
  \mathcal{W} \;=\;
  \operatorname{span}\{\, \underbrace{1}_{\text{constant}};\;
  \underbrace{\xi,\ \eta,\ \zeta}_{\text{linear}};\;
  \underbrace{\xi\zeta,\ \eta\zeta}_{\text{mixed bilinear}} \,\}.
\end{equation*}
A single quadrature point at the barycentre $q_c=(\tfrac13,\tfrac13,0)$ (the
centroid--midpoint rule of degree~1~\citep{felippa2004}) evaluates the element
integrals~\eqref{eq:elemmat}. Constant and linear basis functions are integrated exactly (centroid rule), only the mixed bilinears are subject to quadrature error. To check for potential velocity modes that are missed by the 1-point quadrature rule, we define the element energy
\begin{align}
  E_e(\bm{u}) \;=\; V_e\,
  \eta(q_c)\Big[ 2\,|\bm{\varepsilon}(\bm{u})(q_c)|^2
  - \tfrac{2}{3}\,\big(\mathrm{div}\,\bm{u}(q_c)\big)^2 \Big]\nonumber \\
  \qquad V_e = w_q\,|\det J(q_c)|,
  \label{eq:oneptenergy}
\end{align}
with quadrature weight $w_q$ and the Jacobian determinant of the reference
map evaluated at $q_c$. A velocity that is nonzero, but results in a zero element energy, and is therefore invisible to the integral, is called \textit{hourglass mode} in the finite-element literature~\citep{kosloff1978,flanagan1981}.

By \eqref{eq:oneptenergy}, a nonzero velocity is an hourglass mode exactly if its gradient vanishes
at $q_c$. Writing $u = a\xi + b\eta + c\,\zeta + d\,\xi\zeta +
e\,\eta\zeta \in \mathcal{W}$ per velocity component (omitting constant terms), the condition
$\nabla u(q_c) = (a + d\zeta,\ b + e\zeta,\ c + d\xi + e\eta)\big|_{q_c} = 0$
gives $a = b = 0$, $c = -\tfrac13(d+e)$: a two-dimensional null space
per component, spanned by
\begin{equation}
  w_1 = (\xi - \tfrac13)\,\zeta, \qquad
  w_2 = (\eta - \tfrac13)\,\zeta.
  \label{eq:hourglass}
\end{equation}
These are the hourglass modes of the one-point wedge: scalar amplitude
profiles carried by a fixed velocity direction, e.g. $\bm{u} = w_1\,
\bm{e}_\xi$.
With three velocity components the element stiffness has $2 \times 3 =
6$ spurious zero-energy modes.

\begin{figure}[t]
  \centering
  \includegraphics[width=\linewidth]{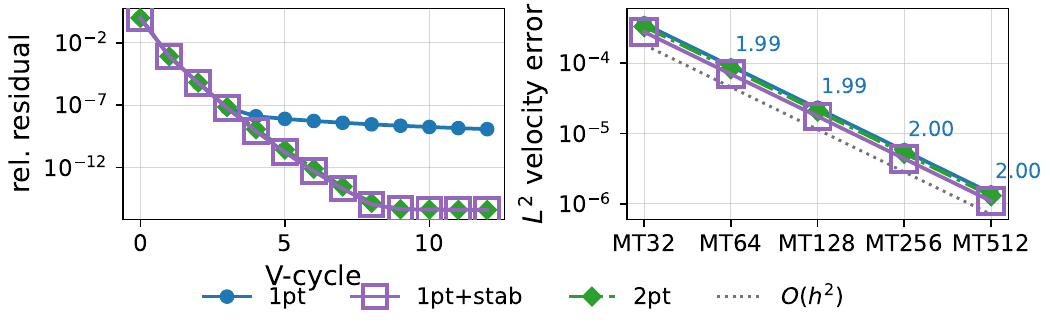}
  \caption{Hourglass control for a smooth analytical solution of the
  viscous diffusion block of~\eqref{eq:stokes} on the spherical shell:
  stand-alone V-cycle residual on the MT256 mesh (169\,M nodes, left)
  and $L^2$ velocity discretization error with observed convergence
  orders over model size MT32--MT512 (right), for one-point quadrature,
  one-point with hourglass control~\eqref{eq:hgstab}, and the
  two-radial-point rule. Both remedies recover textbook multigrid
  convergence (their residual curves coincide, left) at unchanged,
  optimal-order discretization error.}
  \label{fig:hourglass-stab}
\end{figure}

Fortunately, on the whole assembled shell the local modes do not persist, because such local modes can not be stacked together consistently on the whole grid. Thereby, the discretization converges at the optimal rate
(\Cref{fig:hourglass-stab}, right) even with 1 quadrature point. This is contrary to hourglass modes on hexes which persist globally in a checkerboard pattern~\citep{flanagan1981}. They do, however, degrade stand-alone geometric multigrid, demonstrated by
solving the viscous diffusion block of~\eqref{eq:stokes} in \Cref{fig:hourglass-stab}, left.
Both a second radial quadrature point and hourglass
control~\citep{flanagan1981} restore convergence. For the latter, the
element matrix is augmented by the rank-6 correction
\begin{equation}
  A_e^{\mathrm{stab}} \;=\; A_e \;+\; \varepsilon
  \sum_{m=1}^{2}\sum_{d=1}^{3}
  (\bm{w}_m \otimes \bm{e}_d)(\bm{w}_m \otimes \bm{e}_d)^\top,
  \label{eq:hgstab}
\end{equation}
where $\bm{w}_m \in \mathbb{R}^6$ is the vector of nodal values of
$w_m$, its coefficient vector in the nodal basis~\eqref{eq:wedgebasis}, and $\varepsilon$ is a small stabilization parameter scaled with the
local stiffness. Fused into the gather/scatter phases
of the matrix-free kernel (\Cref{sec:kernel}), it adds a few fused
multiply--adds per element and no memory traffic.

\section{Matrix-free Assembly Kernel}
\label{sec:kernel}
Element-local assembly is encapsulated in a Kokkos functor (\texttt{KOKKOS\_FUNCTION operator()}) that is dispatched over the 4D index space via \texttt{Kokkos::parallel\_for}. The same functor code runs on CPU and GPU backends without modification. Its members are the global source and destination vectors \texttt{src\_} and \texttt{dst\_}, held as the structure-of-arrays grid data of \Cref{lst:kokkos-view}, and the per-wedge DoF count \texttt{nwdofs\_}. The host-side \texttt{apply()}, the matrix-free matrix-vector multiplication, dispatches the functor over the 4D index space $(sd, x, y, r)$ of subdomain and lateral and radial cell indices, fences, and exchanges halos, so that contributions to nodes on subdomain boundaries are summed across ranks. \Cref{lst:kernel} shows the overall structure. 

For each of the two wedges $e$, the functor evaluates \eqref{eq:elemmat} directly against the gathered DoFs $\bm{x}_e$ without ever forming $A_e$. At each quadrature point $q$ it contracts the source into the discrete strain rate
\begin{equation}
  \bm{\varepsilon}_q \;=\; \sum_{j=1}^{18} \bm{\varepsilon}(\bm{\varphi}_j)\big|_q\; (\bm{x}_e)_j,
  \label{eq:epsq}
\end{equation}
and accumulates its contraction with every test gradient,
\begin{equation}
  (\bm{y}_e)_i \;=\; \sum_q w_q\,|\det J_q|\,\eta_q 
  \times \Bigl[\, 2\,\bm{\varepsilon}(\bm{\varphi}_i)\big|_q : \bm{\varepsilon}_q
  \;-\; \tfrac{2}{3}\,\mathrm{div}(\bm{\varphi}_i)\big|_q\, \mathrm{tr}\,\bm{\varepsilon}_q \,\Bigr],
  \label{eq:localapply}
\end{equation}
for $i = 1,\dots,18$. The local result $\bm{y}_e$ is scattered into the global destination by $I_e^\top$ using atomic addition.

\begin{lstlisting}[
  caption={Structure of the viscous operator: the element-local assembly is a single Kokkos functor dispatched over the 4D index space $(sd, x, y, r)$, one thread per hexahedral cell.},
  label={lst:kernel},
  float=t,
  basicstyle=\ttfamily\scriptsize,
]
struct ViscousOperator_Textbook {
 Grid4DDataVec<real,3> src_, dst_;
 constexpr int nwdofs_ = 6;   // nodes per wedge

 void apply() {
  Kokkos::parallel_for(
   MDRangePolicy<Rank<4>>({0,..}, {n_sd,nx,ny,nr}), *this);
  Kokkos::fence();
  mpi_halo_exchange(dst_);
 }

 KOKKOS_FUNCTION
 void operator()(int sd, int x, int y, int r) const {
  ...
 }
};
\end{lstlisting}

A range of GPU-oriented optimisations turn this textbook baseline into a bandwidth-bound implementation that reaches a substantial fraction of peak performance across NVIDIA, AMD, and Intel GPUs. We defer the description to the companion performance paper~\citep{boehm2026performance}. 

\section{Time Discretization}
\label{sec:timediscretization}

\projectname{} provides several schemes for the energy equation~\eqref{eq:energy}: the classical streamline-upwind Petrov--Galerkin (SUPG) stabilisation~\citep{brooks1982supg}, the entropy-viscosity method~\citep{guermond2011entropy} as used in ASPECT~\citep{kronbichler2012}, and, as the main scheme, the modified method of characteristics (MMOC)~\citep{douglas1982mmoc,kohl2022mmoc}.

A characteristic is the trajectory $\mathbf{X}(t)$ of a fluid particle, $\mathrm{d}\mathbf{X}/\mathrm{d}t = \mathbf{u}(\mathbf{X},t)$. Following such a particle, the chain rule gives
\begin{equation}
  \frac{\mathrm{d}}{\mathrm{d}t}\,T\bigl(\mathbf{X}(t),t\bigr) \;=\; \frac{\partial T}{\partial t} + \nabla T\cdot\frac{\mathrm{d}\mathbf{X}}{\mathrm{d}t} \;=\; \frac{\partial T}{\partial t} + \mathbf{u}\cdot\nabla T,
  \label{eq:characteristic}
\end{equation}
i.e.\ the advective derivative in~\eqref{eq:energy} is the ordinary time derivative of the temperature seen by the moving particle. The MMOC exploits this by following the flow backwards: the characteristic that arrives at a node $\mathbf{x}_i$ at $t^{n+1}$ is traced back over one timestep to its departure point
\begin{equation}
  \hat{\mathbf{x}}_i \;=\; \mathbf{x}_i - \int_{t^{n}}^{t^{n+1}} \mathbf{u}\bigl(\mathbf{X}(t),t\bigr)\,\mathrm{d}t,
  \label{eq:departure}
\end{equation}
integrated with an explicit Runge--Kutta method, and the old temperature is interpolated there, $\hat T^{n}_i = T^{n}(\hat{\mathbf{x}}_i)$. Replacing the advective derivative by the difference quotient along the characteristic and treating diffusion and sources implicitly on the fixed grid gives
\begin{equation}
  \frac{T^{n+1} - \hat T^{n}}{\Delta t} \;-\; \frac{1}{\bar\rho\,\bar c_p}\,\nabla\cdot\bigl(k\,\nabla T^{n+1}\bigr) \;=\; Q,
  \label{eq:mmoc}
\end{equation}
with $Q$ the right-hand side of~\eqref{eq:energy}. Each timestep, thus, splits into an advection step, i.e.~the evaluation of $T^{n}$ at all departure points, and a diffusion step, the solution of~\eqref{eq:mmoc} with matrix-free preconditioned CG. As advection is carried along the flow rather than through a stencil, the scheme is not CFL-bound for stability, admits considerably larger timesteps and introduces little numerical diffusion.
In \projectname{}, each particle is fully virtual (holding no information) and traced back to its departure point by a dedicated thread in a Kokkos kernel. At its departure point, the temperature is interpolated using linear, or higher order interpolation, with cross GPU boundaries handled by enlarged boundary layers. A more detailed analysis of the scheme can be found in~\citep{kohl2022mmoc}.
\section{Solver Structure}
\label{sec:agca}

The Stokes saddle-point matrix is preconditioned by the standard block-triangular Schur-complement approach~\citep{elman2005} widely-used across mantle convection codes~\citep{maymoresi2008,burstedde2013,kronbichler2012}. Writing the discretised system as $\bigl(\begin{smallmatrix} A & B^T \\ B & 0 \end{smallmatrix}\bigr)\bigl(\begin{smallmatrix}\mathbf{u}\\p\end{smallmatrix}\bigr) = \bigl(\begin{smallmatrix}\mathbf{f}\\\mathbf{g}\end{smallmatrix}\bigr)$, the preconditioner reads
\begin{equation}
  \mathcal{P}^{-1} =
  \begin{pmatrix} A^{-1} & -A^{-1} B^{T}\, \hat{S}^{-1} \\ 0 & \hat{S}^{-1} \end{pmatrix},
  \label{eq:block-precond}
\end{equation}
with $\hat{S}^{-1}$ approximating the inverse of the Schur complement $S = -B A^{-1} B^{T}$ by the scaled diagonal of the inverse-viscosity-weighted pressure mass matrix,
\begin{equation}
  \hat{S}^{-1} = -\operatorname{diag}\bigl(M_{1/\eta}\bigr)^{-1},
  \qquad
  \bigl(M_{1/\eta}\bigr)_{ij} = \int_{\Omega} \frac{1}{\eta}\, \psi_i\, \psi_j \,\mathrm{d}\Omega,
  \label{eq:pmass}
\end{equation}
with the pressure basis functions $\psi_i$: for moderate viscosity contrasts, $M_{1/\eta}$ is spectrally equivalent to the (negative) Schur complement for variable viscosity~\citep{elman2005}.
The velocity-block inverse $A^{-1}$ is realised approximately by a matrix-free geometric multigrid V-cycle with Chebyshev smoothing~\citep{elman2005,rudi2015}, which is the dominant cost of every FGMRES~\citep{saad1993} iteration in the outer solver.

The V-cycle applies two pre- and two post-smoothing steps of a second-degree Chebyshev polynomial smoother on every level and solves the coarsest-grid system with conjugate gradients.
The radial viscosity structure of the mantle is what limits the performance of these solver components: realistic profiles vary by orders of magnitude over a few hundred kilometres, with the steepest gradients at the base of the lithosphere and across the asthenosphere~\citep{lin2022viscosity,stotz2017viscosity}. Such coefficient contrasts degrade the smoother and coarse-grid correction of the velocity multigrid as well as the Schur approximation~\eqref{eq:pmass}. We therefore consider two radial profiles of different severity: the profile of Lin et al.~\citeyear{lin2022viscosity}, spanning three orders of magnitude ($4.7\times10^{20}$ -- $5\times10^{23}$\,Pa\,s), and the profile of Stotz et al.~\citeyear{stotz2017viscosity}, spanning four orders ($5.8\times10^{19}$ -- $7\times10^{23}$\,Pa\,s) with a pronounced low-viscosity asthenosphere.

\begin{figure}[t]
  \centering
  \begin{minipage}[c]{0.40\textwidth}
    \centering
    \includegraphics[width=\linewidth]{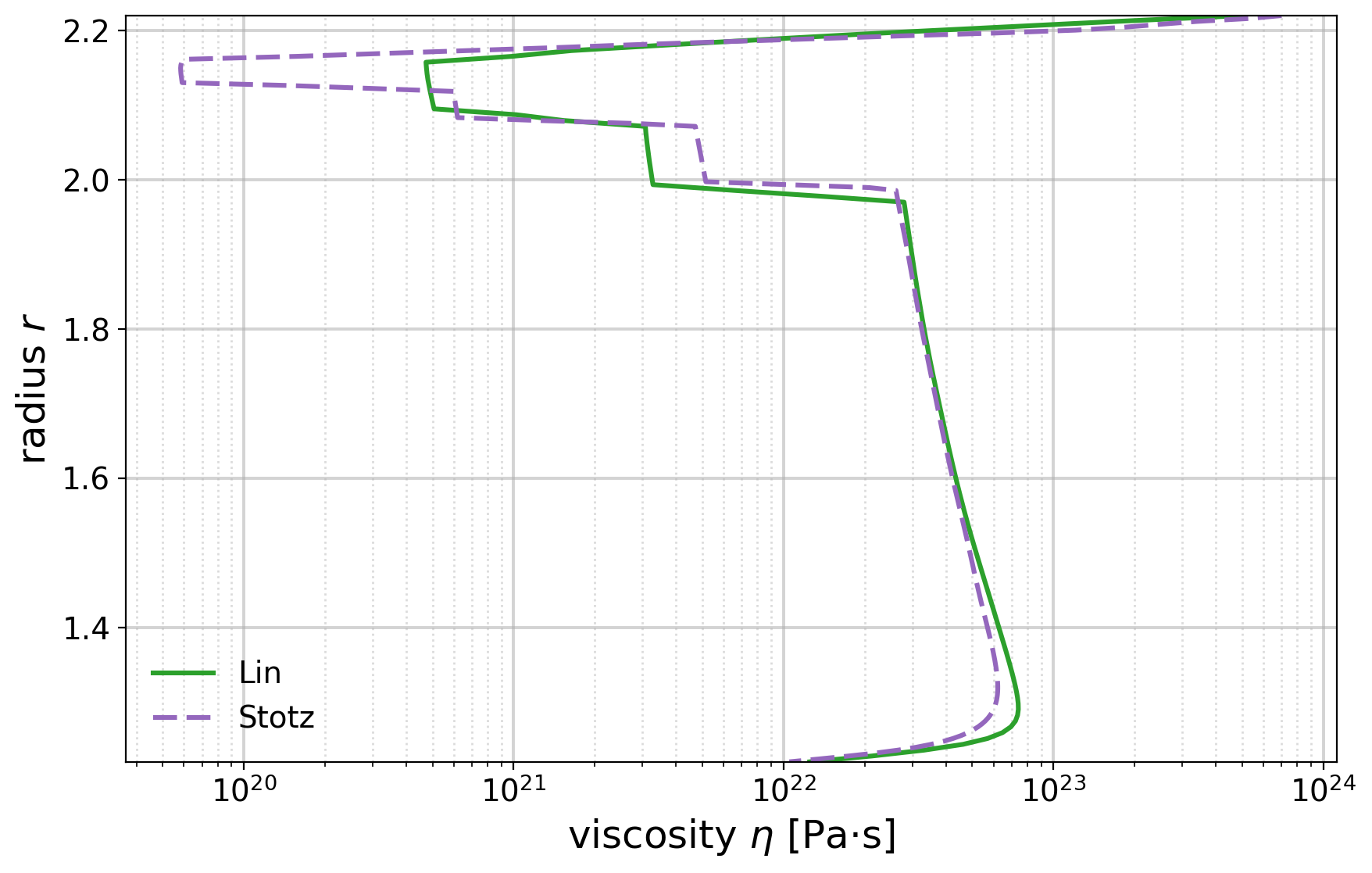}
  \end{minipage}\hfill
  \begin{minipage}[c]{0.28\textwidth}
    \centering
    \footnotesize
    \setlength{\tabcolsep}{5pt}
    \renewcommand{\arraystretch}{1.05}
    \begin{tabular}{c cc}
    \toprule
    MT     & Lin & Stotz \\
    \midrule
    MT32   & 41 & 49 \\
    MT64   & 32 & 45 \\
    MT128  & 29 & 37 \\
    MT256  & 26 & 33 \\
    MT512  & 27 & 33 \\
    MT1024 & 31 & 40 \\
    MT2048 & 31 & 42 \\
    \bottomrule
    \end{tabular}\\[3pt]
    \footnotesize\textit{FGMRES iters to $10^{-6}$ relative residual.}
  \end{minipage}\hfill
  \begin{minipage}[c]{0.28\textwidth}
    \centering
    \footnotesize
    \setlength{\tabcolsep}{5pt}
    \renewcommand{\arraystretch}{1.05}
    \begin{tabular}{c cc}
    \toprule
    MT     & Lin & Stotz \\
    \midrule
    MT32   & $3.7\!\times\!10^{-4}$ & $3.5\!\times\!10^{-4}$ \\
    MT64   & $2.9\!\times\!10^{-4}$ & $3.0\!\times\!10^{-4}$ \\
    MT128  & $2.4\!\times\!10^{-4}$ & $3.1\!\times\!10^{-4}$ \\
    MT256  & $1.63\!\times\!10^{-4}$ & $2.37\!\times\!10^{-4}$ \\
    MT512  & $1.12\!\times\!10^{-4}$ & $2.25\!\times\!10^{-4}$ \\
    MT1024 & $6.88\!\times\!10^{-4}$ & $1.57\!\times\!10^{-3}$ \\
    MT2048 & $7.09\!\times\!10^{-4}$ & $1.52\!\times\!10^{-3}$ \\
    \bottomrule
    \end{tabular}\\[3pt]
    \footnotesize\textit{Relative residual after 10 FGMRES iters}
  \end{minipage}
  \caption{Left: radial viscosity profiles of \citet{lin2022viscosity} and \citet{stotz2017viscosity} used for the convergence study. Centre: full Stokes FGMRES iterations to reduce the relative residual by six orders of magnitude. Right: relative residual reached after a fixed budget of 10 outer FGMRES iterations.}
  \label{fig:visc-profiles-convergence}
\end{figure}

\Cref{fig:visc-profiles-convergence} documents convergence of the full Stokes solve for both profiles. The left table reports the number of preconditioned FGMRES iterations needed to reduce the relative residual by six orders of magnitude; the right table reports the relative residual actually achieved after a fixed budget of 10 outer FGMRES iterations, which is a realistic configuration in production runs. The results confirm that the multigrid preconditioner delivers mesh-independent convergence with real-world viscosity profiles. There is a light drift for the large models, which is expected for the light-weight pressure-mass preconditioner (in contrast to a stronger BFBT-type preconditioner~\citep{rudi2017bfbt,Burkhart:2026:GEM}), but convergence is satisfactory. The cell-to-cell coefficient contrast seen by the smoother and the coarse grids decreases with model size, as the smooth profile is resolved more finely, so the solve initially gets easier as the problem size grows. 

For even sharper contrasts, \projectname{} possesses a GPU-accelerated, wedge-based implementation of adaptive Galerkin coarsening~\citep{boehm2026agca}. There, assembled Galerkin coarse-grid operators are built element-wise only where the viscosity gradient exceeds a threshold, typically a thin band along the boundary layers, while the cheap matrix-free re-discretisation is kept everywhere else, restoring robust multigrid convergence under strong coefficient contrasts at minimal memory overhead.

\section{Verification Benchmarks}
\label{sec:validation}

\begin{figure}[t]
  \centering
  \includegraphics[width=0.92\textwidth]{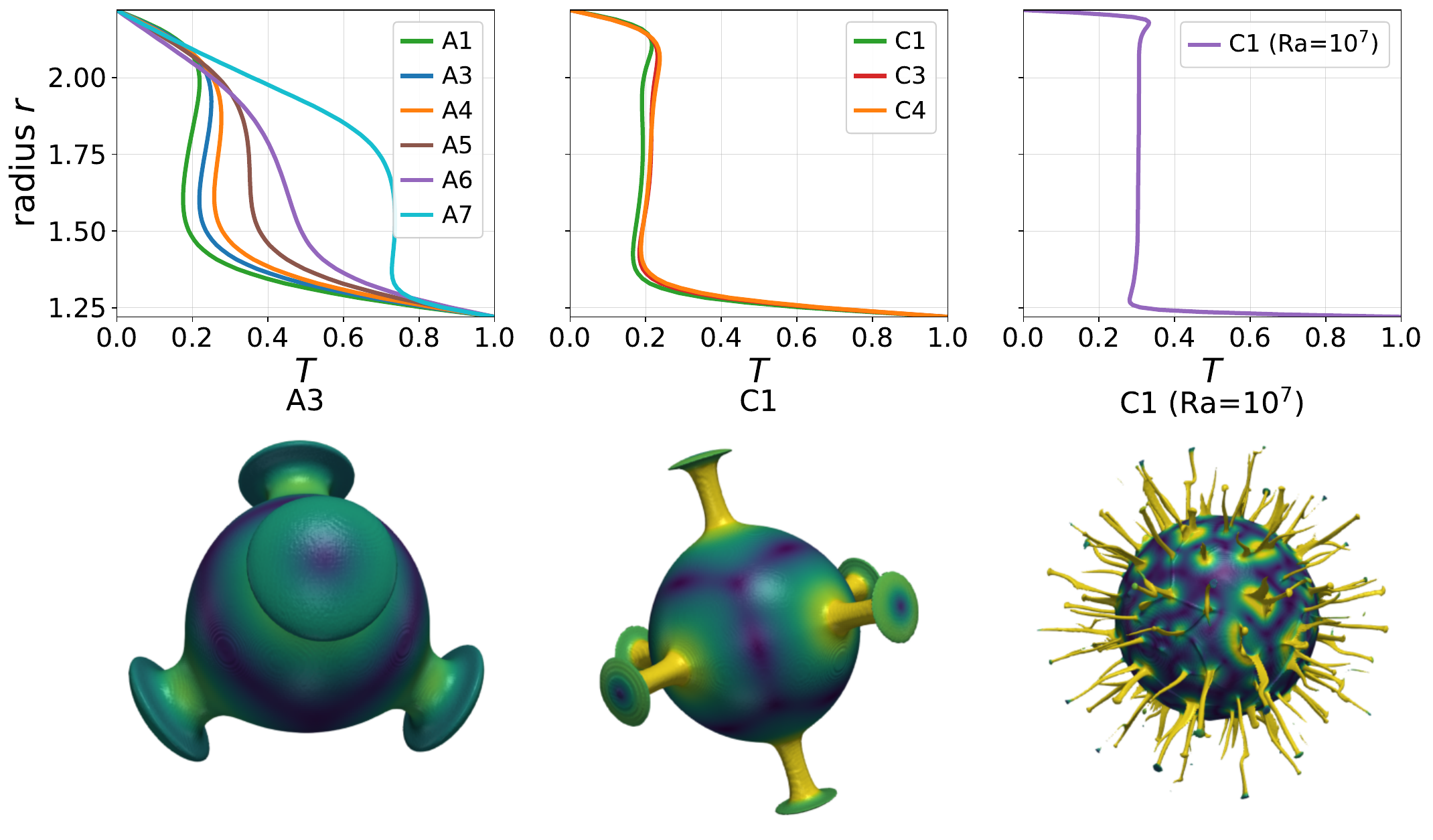}
  \caption{Verification benchmarks: steady-state radial temperature profiles $\langle T\rangle(r)$ (top row) and, beneath each, the steady-state $T = 0.5$ isosurface of the named case, colored by velocity magnitude at the same time. These benchmarks were run using MMOC for the energy equation. Entropy viscosity gave the same or very similar results in terms of final isosurface, Nusselt number and radial profile. }
  \label{fig:mc-validation}
\end{figure}

We verify the coupled Stokes-energy framework against a set of spherical shell convection solutions, explored early on for steady flow (e.g. \citet{Bercovici_Schubert_Glatzmaier_Zebib_1989, Ratcliff_etal_1996, zhong2000, Richards_etal_2001}), extended in the mantle convection benchmark suite of \citet{zhong2008}, and compiled and reused in the recent code intercomparison of \citet{euen2023} and the HyTeG validation of \citet{ilangovan2026}. We pick cases that span the relevant parameter range, from mild low-Rayleigh tetrahedral convection (A1--A7, increasing viscosity contrast) up to vigorous high-Rayleigh cubic convection at $\mathrm{Ra}=10^{5}-10^{7}$ with C1--C4:

\begin{center}
\footnotesize
\setlength{\tabcolsep}{4pt}
\begin{tabular}{lccl}
\toprule
case & $\mathrm{Ra}$ & $\Delta\eta$ & $T$ perturb \\
\midrule
A series         & $7\times 10^{3}$ & $\{1,\,20,\,10^{2},\,10^{3},\,10^{4},\,10^{5}\}$ & $Y_{3}^{\,2}$ \\
C series         & $1\times 10^{5}$ & $\{1,\,30,\,100\}$                     & $Y_{4}^{\,0} + \tfrac{5}{7}Y_{4}^{\,4}$ \\
C1 (Ra=$10^{7}$) & $1\times 10^{7}$ & $1$                                    & $Y_{4}^{\,0} + \tfrac{5}{7}Y_{4}^{\,4}$ \\
\bottomrule
\end{tabular}
\end{center}
We run all cases with an MT256 resolution, use the non-dimensional shell $r\in[1.22,2.22]$, free-slip boundary conditions on the top and bottom surface, isothermal Dirichlet temperature conditions ($T_\mathrm{cmb}=1$, $T_\mathrm{surf}=0$), and the Frank--Kamenetskii viscosity law $\eta(T) = \exp\bigl(-E(T-\tfrac{1}{2})\bigr)$, with the viscosity contrast $\Delta\eta = \exp(E)$ between top and bottom matching the value tabulated by \citet{zhong2008}. Each case is advanced until the radially averaged temperature profile and the surface Nusselt number $\mathrm{Nu}_\mathrm{top}$ stabilise; we then report the steady-state $\langle T\rangle(r)$ profile (\Cref{fig:mc-validation}) and the steady-state $\mathrm{Nu}_\mathrm{top}$. For A3, C1, and C3 the computed $\mathrm{Nu}_\mathrm{top}$ lies inside the reference interval reported by the intercomparison studies; for the temperature-dependent-viscosity cases A4 and A5 we compare against the CitcomS reference values of \citet{zhong2008}; and for the high-Ra C1 ($\mathrm{Ra}=10^{7}$) case we compare against the recent isoviscous spherical-shell result of \citet{garridotomasini2025}. Across the board, the Nusselt number lies in the expected interval or is very close to the reference value given by literature.

\begin{table}[t]
\centering
\footnotesize
\setlength{\tabcolsep}{3.5pt}%
\caption{Steady-state $\mathrm{Nu}_\mathrm{top}$ vs.\ the reference intervals from \citet{euen2023, ilangovan2026} (A3, C1, C3), the CitcomS reference values of \citet{zhong2008} (A1, A4--A7, C4), and the isoviscous Ra=$10^{7}$ value of \citet{garridotomasini2025} (C1$^\star$ denotes C1 at Ra=$10^{7}$).}
\label{tab:mc-nu}
\begin{tabular}{l cccccc cccc}
\toprule
case & A1 & A3 & A4 & A5 & A6 & A7 & C1 & C3 & C4 & C1$^\star$ \\
\midrule
$\mathrm{Nu}_\mathrm{top}$ (\projectname{})
 & $3.49$ & $3.162$ & $2.99$ & $2.54$ & $2.055$ & $2.716$
 & $7.714$ & $6.778$ & $6.496$ & $31.1$ \\
reference
 & $3.5126$ & $[3.140, 3.190]$ & $2.9354$ & $2.5468$ & $2.0414$ & $2.7382$
 & $[7.370, 7.810]$ & $[6.500, 6.790]$ & $6.4800$ & $31.91$ \\
\bottomrule
\end{tabular}
\end{table}

\section{Strong and Weak Scaling of TERRA-NG}
\label{sec:fullsim}

The simulations in this section advance global mantle-circulation modelling to resolutions that, to the best of our knowledge, have not been reported before: coupled Stokes--energy timestepping with compressibility and variable viscosity beyond MT512, which is the upper end of published global models, up to MT4096 and, in a low-memory mode, MT8192.
Furthermore, we show portability of the code to five supercomputers spanning three GPU vendors.
\Cref{tab:sites} lists the five host systems on which the simulations are run. We included 2 NVIDIA, 2 AMD and 1 Intel machine for a representative vendor study. We used benchmarking allocations through the EuroHPC Joint Undertaking for most machines, except LUMI-G, where we computed using an extreme-scale access.

\begin{table*}[t]
\centering
\setlength{\tabcolsep}{10pt}
\caption{Host supercomputers of the feasibility study, with operating compute centre, accelerator-partition node count, per-node GPU architecture, and aggregate main memory of the GPU partition (host DRAM $+$ per-accelerator HBM). All systems carry 512\,GB of node-level main memory (DDR4/DDR5 on JUWELS Booster, MareNostrum 5, LUMI-G and SuperMUC-NG; unified HBM3 on the Hunter MI300A APU nodes).}
\label{tab:sites}
\begin{tabular}{lllrlr}
\toprule
system & vendor & computing centre & nodes & Architecture & main memory \\
\midrule
JUWELS Booster      & NVIDIA & JSC, J\"ulich, DE      & 936     & A100 SXM4      & 480\,TB  \\
MareNostrum 5 ACC   & NVIDIA & BSC, Barcelona, ES     & 1{,}120 & H100 SXM5      & 573\,TB  \\
LUMI-G              & AMD    & CSC, Kajaani, FI       & 2{,}978 & MI250X         & 1.5\,PB  \\
Hunter              & AMD    & HLRS, Stuttgart, DE    & 188     & 4 MI300A APU   & 96\,TB   \\
SuperMUC-NG Phase 2 & Intel  & LRZ, Munich, DE        & 234     & PVC            & 120\,TB  \\
\bottomrule
\end{tabular}
\end{table*}

We evaluate the feasibility of running coupled Stokes--energy mantle-convection simulations up to a given spatial resolution on each of the considered GPU supercomputers, and identify the optimal node configuration for every resolution through a per-MT-level strong-scaling sweep. Each timestep solves the Stokes saddle-point system with 10 preconditioned FGMRES iterations and the energy equation with 50 preconditioned FGMRES iterations.
We run 10 timesteps with noslip-freeslip boundary conditions and the Lin viscosity profile.
The low-memory mode sets the FGMRES restarts to 5 instead of 10 and stores the Krylov basis vectors in single precision, reducing the memory the Krylov basis requires.

\begin{figure*}[h]
\centering
\includegraphics[width=\textwidth]{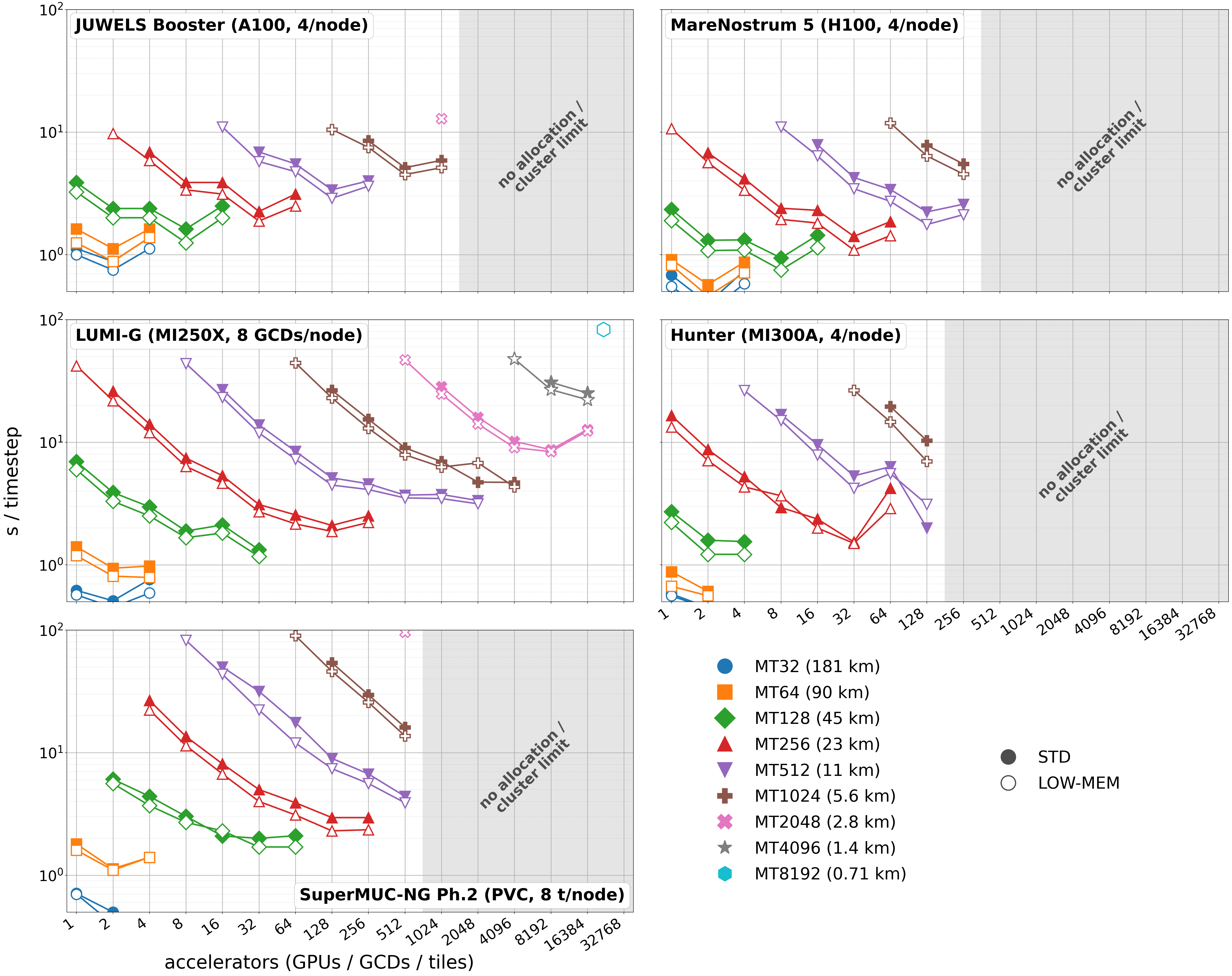}
\caption{Strong scaling of the coupled Stokes--energy timestep across five GPU supercomputers, with one line per MT Model and corresponding grid spacing (\Cref{tab:mt-resolution}).The scaling results were run using entropy viscosity for the energy equation, because MMOC was implemented after the accesses to some of these machines.}
\label{fig:cross-strong}
\end{figure*}

\Cref{fig:cross-strong} plots the average time per simulation timestep over the MT configurations. Each line is a strong-scaling sweep: the model size is held fixed and spread over more and more GPUs, so the curve falls until the per-device work no longer pays for the added communication, and speed-ups from more devices diminish. The figure also carries a weak-scaling reading across its lines. Every step to the next MT level multiplies the number of unknowns by eight, four times laterally and twice radially, so pairing that step with an eight-fold increase in devices keeps the work per device constant. Read along that diagonal, the curve starts line up on a roughly straight weak-scaling line, especially visible for LUMI-G: MT256 on $2$ GCDs, MT512 on $16$, MT1024 on $128$, MT2048 on $1024$ and MT4096 on $8192$ all hold $126$\,M DoFs per GCD, and the time per timestep grows from $26.2$ to $30.9$\,s across that chain, an $18\,\%$ increase for a $4096$-fold growth in both device count and problem size.

We observe that simulations up to MT128 run with $<$ 10 seconds per timestep on all systems. Resolutions of MT256 -- MT1024, at and beyond the upper end of published global mantle-convection simulations, run routinely on most systems with $<$ 20 seconds per timestep. TERRA-NG allows running MT4096 ($1.4 \times 10^{12}$ DoFs) reaching the 1-km-per-gridpoint resolution with the extreme-scale access on LUMI-G at $ \sim$ 30 seconds per timestep. On the full machine, an MT8192 with $1.1 \times 10^{13}$ DoFs is feasible in low-memory-mode which reaches sub-1-km-resolution. Already MT1024 exceeds, to the best of the authors' knowledge, any published global mantle-convection simulation with compressibility and variable viscosity, MT4096 and MT8192 push this frontier by one to two further orders of magnitude in DoFs.

\section{End-to-end Simulation}
\label{sec:e2e}

To demonstrate \projectname{} as a production tool, we run an end-to-end mantle-circulation model at MT1024 on 64 nodes
(512 PVC tiles, one rank per tile) of SuperMUC-NG Phase~2: at $5.6$\,km radial and ${\sim}7$\,km lateral spacing, i.e.\
$1.7\times10^{10}$ velocity--pressure DoF per Stokes solve, and $2.2\times10^{10}$ DoF in total once the temperature
field is counted. The model is TALA-compressible with a $\bar\rho$-weighted buoyancy on an adiabatic
reference state, free-slip at both boundaries, and fixed temperatures of
$4200$\,K at the CMB and $300$\,K at the surface. The viscosity is the radial profile of Lin et al.\ (2022) multiplied by a
Frank--Kamenetskii temperature dependence of contrast $\Delta\eta_T=100$. The reference viscosity is
$3.4\times10^{20}$\,Pa\,s, the minimum of that radial profile, which puts the Rayleigh number formed with it at
$\text{Ra}\approx5\times10^{8}$. The initial condition is a conductive
profile perturbed by a broadband spherical-harmonic field of degrees
$\ell=8$--$96$ (amplitude $0.05$), so that no low-degree symmetry is imposed.

\begin{figure*}[t]
  \centering
  \begin{minipage}[c]{0.54\textwidth}
    \centering
    \includegraphics[width=\linewidth]{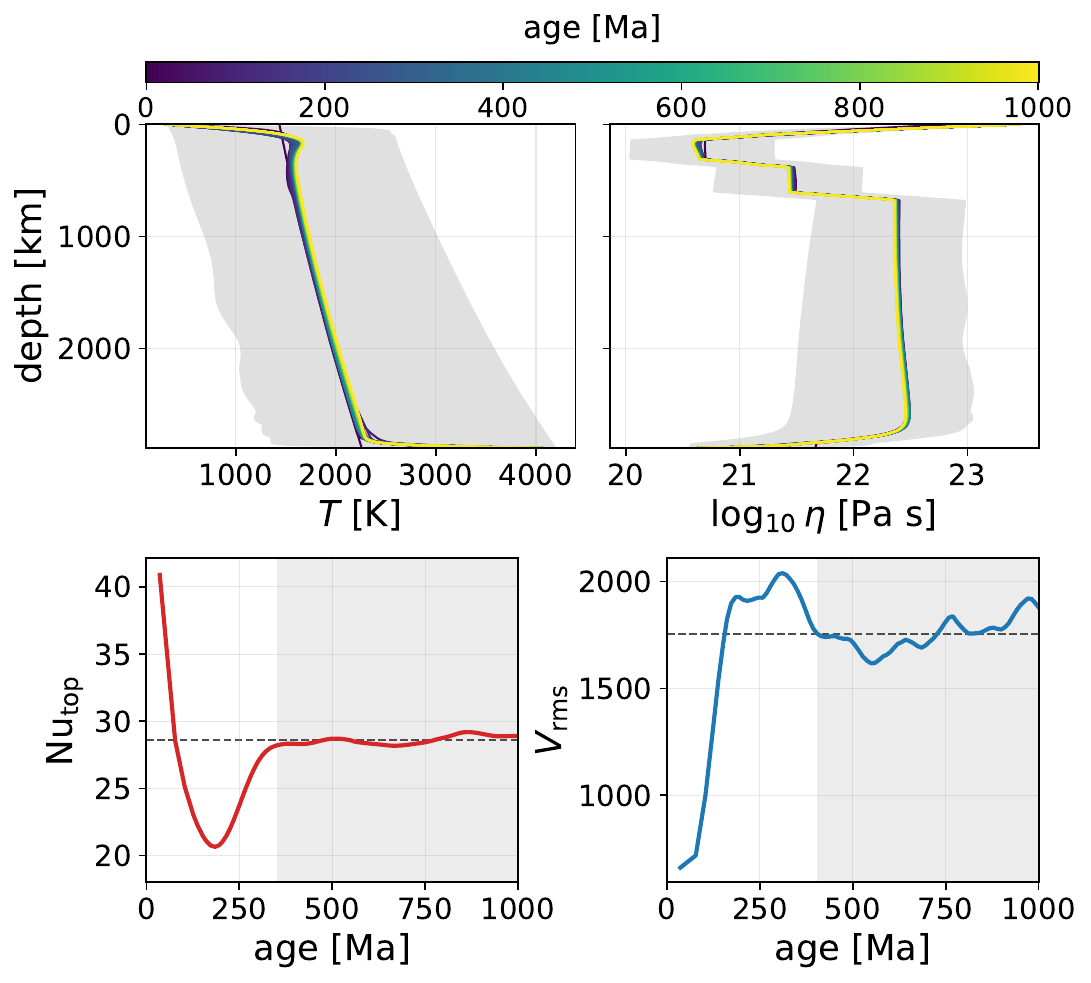}
  \end{minipage}\hfill
  \begin{minipage}[c]{0.44\textwidth}
    \centering
    \includegraphics[width=0.86\linewidth]{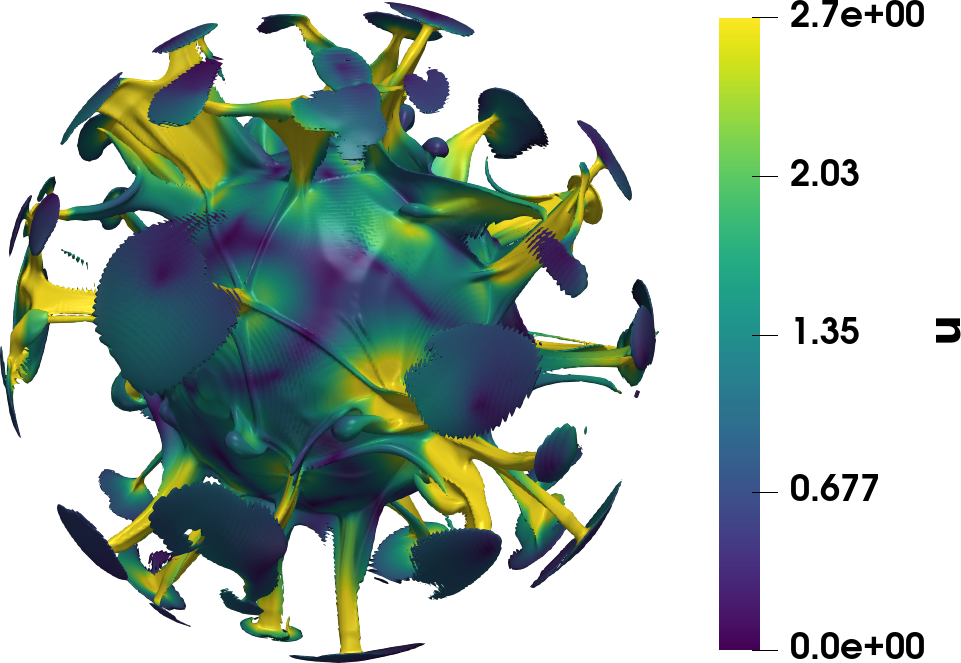}\\[2pt]
    \includegraphics[width=\linewidth]{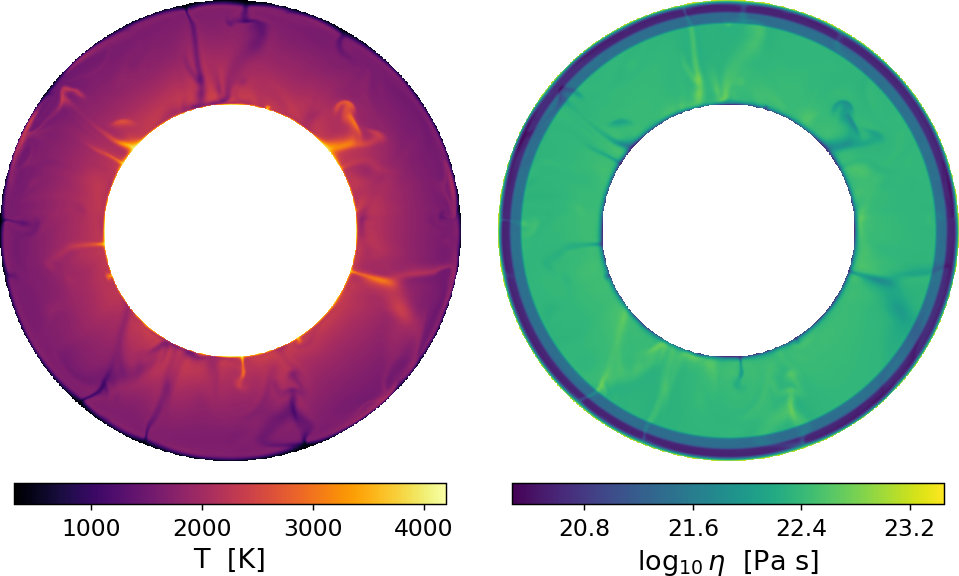}
  \end{minipage}
  \caption{End-to-end MT1024 simulation. Left block, top row: radial-mean
  temperature and viscosity every $100$ timesteps, coloured by model age, over
  the lateral min--max range of the final state in grey. The imposed Lin
  staircase is overlaid by the Frank--Kamenetskii response of the lid,
  asthenosphere and basal layer to the thermal boundary layers. Bottom row:
  surface Nusselt number and root-mean-square velocity against model age, with
  the steady window shaded and its mean dashed. Right, top: isosurface of the
  temperature deviation from the radial mean at $+400$\,K, coloured by velocity
  magnitude. Right, bottom: meridional cross-cut of temperature and viscosity
  through the same state at $t=1.00$\,Gyr. This end-to-end simulation was run with MMOC for the energy equation.}
  \label{fig:e2e}
\end{figure*}

Each timestep solves the Stokes system with 10 FGMRES iterations on the
multigrid-preconditioned velocity block and transports the temperature with the
MMOC semi-Lagrangian scheme at a Courant number of $2.7$, whose diffusion step takes at most 20 Krylov iterations at a
$10^{-3}$ relative tolerance, at ${\sim}19$\,s of wall time per step including output. The
CFL-controlled timestep shrinks from $0.32$\,Ma in the initial layered state to
$0.10$\,Ma once the convective flow is fully developed. $8800$ steps advance the
model over $1.00$\,Gyr across two chained 24-hour allocations, i.e.\ ${\sim}23$\,Ma of
mantle evolution per compute hour. The two diagnostics settle at different times (\Cref{fig:e2e}, bottom left):
the surface Nusselt number is flat at $\mathrm{Nu}_\mathrm{top} \sim 28.6 $
from ${\sim}350$\,Ma onward, whereas the root-mean-square velocity overshoots and only then settles at
$V_\mathrm{rms} \sim 1.8\times10^{3}$. 

\section*{\conclusionname}
Global mantle-circulation modelling has long been resolution-bound.
\projectname{} moves that bound to kilometre-scale grid spacing through a
deliberately narrow design: a single application, one mesh geometry, and discretization tailored towards that geometry rather than a
general finite-element abstraction. That narrow design is what makes the substantial domain-specific optimizations possible, and it is those optimizations, together with the hardware capabilities of modern GPUs, that carry the resolution. Thanks to Kokkos, \projectname{} runs on
NVIDIA, AMD and Intel GPUs, demonstrated on five supercomputers, enabling compressible, variable-viscosity mantle circulation at
\begin{itemize}[nosep,leftmargin=*]
  \item $\sim\!5.6$\,km grid spacing ($22$\,B DoFs) in under $20$ seconds per
        timestep on all five systems,
  \item $\sim\!2.8$\,km and $\sim\!1.4$\,km grid spacing ($177$\,B and
        $1.4 $ T DoFs) on the extreme-scale allocation of LUMI-G,
        the latter at $\sim\!30$ seconds per timestep.
\end{itemize}

The value of these resolutions lies less in the individual model than in
what becomes routine. Mantle-circulation models supply quantitative input to
neighbouring geoscience disciplines, the core--mantle-boundary heat-flux
pattern that drives geodynamo simulations, and the mantle heterogeneity structure against
which seismological, geodetic and geologic observations can be interpreted technically, but
only if they can be produced at sufficient resolution. Timesteps of seconds at kilometre-scale resolution and high model complexity turn such
campaigns from single hero runs into a production workflow, which is where
the results of this paper are meant to lead.
\\
\\
\setcounter{section}{0}
\renewcommand{\thesection}{\thepart.\arabic{section}}

\emph{AI disclosure.}  TERRA-NG contains a subset of code that was developed with Claude Code (Anthropic) as an interactive coding assistant, under the close supervision of the authors, who gave detailed instructions and requirements. All functionality, whether written by humans or with machine assistance, has been thoroughly reviewed and tested by the authors, who take full responsibility for it.
During the preparation of this manuscript, the authors used Claude (Anthropic) to draft and improve the readability and language of the text, while all structure and content were provided by the authors. No technical claims, figures or interpretations are AI-generated. The authors reviewed and edited all content and take full responsibility for it.
\codedataavailability{TERRA-NG is developed openly at \url{https://github.com/mantleconvection/TERRA-NG} under the GNU General Public License v3. The exact version described in this paper, including the configurations and batch scripts under \\ \texttt{apps/mantlecirculation/terrangpaper/} that reproduce all figures and tables, is archived on Zenodo at \\ \url{https://doi.org/10.5281/zenodo.22938125} \citep{boehm2026terrangcode}.}

\authorcontribution{FB designed and implemented the matrix-free wedge
operators, carried out the verification benchmarks and the
scaling campaign, and wrote the manuscript. NK designed the grid and
parallelisation layer, the linear-algebra and solver infrastructure and the
test framework, and contributed to the operator implementation. GR and FB developed
the mantle-circulation application and its diagnostics. PI contributed
finite-element operators and linear forms. PI and FB contributed the MMOC scheme. FR and GR contributed the plate velocity boundary conditions in the mantle-circulation application.
GR, PI, MM, BSAS and HPB provided the
geodynamical problem setting, defined the verification benchmarks and
interpreted the results. HK and UR supervised the HPC design and the
performance work. FB, BSAS, HK, HPB and UR acquired funding and compute resources.
All authors read, revised and approved the manuscript.}

\competinginterests{The authors declare that they have no competing interests.}

\begin{acknowledgements}
The authors gratefully acknowledge the scientific support and HPC resources
provided by the Erlangen National High Performance Computing Center (NHR@FAU)
of the Friedrich--Alexander--Universit\"at Erlangen--N\"urnberg (FAU). The
hardware is partially funded by the German Research Foundation (DFG).
We acknowledge the EuroHPC Joint Undertaking for awarding this project access to
the EuroHPC supercomputer LUMI, hosted by CSC (Finland) and the LUMI consortium
through a EuroHPC Regular Access call (EHPC-BEN-2026B05-001).
We acknowledge the EuroHPC Joint Undertaking for awarding us access to
MareNostrum5 at BSC, Spain (EHPC-BEN-2026B04-050).
The authors gratefully acknowledge the Gauss Centre for Supercomputing e.V.
(\url{www.gauss-centre.eu}) for funding this project by providing computing time
on the GCS Supercomputer SuperMUC-NG at Leibniz Supercomputing Centre
(\url{www.lrz.de}) Project pn25xe.
The simulations were performed on the national supercomputer HPE Cray EX4000
Hunter at the High Performance Computing Center Stuttgart (HLRS).
\end{acknowledgements}

\bibliographystyle{copernicus}
\bibliography{references}

\end{document}